\documentclass{elsart}
\usepackage{graphicx}
\usepackage{amsmath}
\usepackage{amssymb}
\usepackage{subfigure}
\journal{Physica A}

\begin{document}

\begin{frontmatter}

\title{Non-extensive Behavior of a Stock Market Index at Microscopic Time Scales}
\author[address1]{A.A.G.Cortines,}
\author[address1]{R. Riera\thanksref{thank1}}

\address[address1]{Departamento de F\'{\i}sica, Pontif\'{\i}cia Universidade
Cat\'olica, C.P. 38071-970, Rio de Janeiro, RJ, Brazil}

\thanks[thank1]{Corresponding author: Dr.Rosane Riera Freire, Departamento de F\'{\i}sica, Rua Marqu\^es de S\~ao Vicente 225, G\'avea 22453-900, Rio de Janeiro, Rio de Janeiro, Brasil. Tel: (55) (021) 3114-1263; fax: (55) (021) 3114-1271
      E-mail addresses: rrif@fis.puc-rio.br (R. Riera), cortines@fis.puc-rio.br (A.A.G. Cortines)}

\ead{rrif@fis.puc-rio.br}      
      
\begin{abstract}
This paper presents an empirical investigation of the intraday Brazilian stock market price fluctuations, 
considering q-Gaussian distributions that emerge from a non-extensive statistical mechanics. Our results show that,
when returns are measured over intervals less than one hour, the empirical distributions are well fitted by q-Gaussians
with exponential damped tails. Scaling behavior is also observed for these microscopic time intervals. We find that the
time evolution of the distributions is according to a super diffusive q-Gaussian stationary process within
a nonlinear Fokker-Planck equation. This regime breaks down due to the exponential fall-off of
the tails, which in turn, governs the transient dynamics to the long-term macroscopic Gaussian regime. Our results
suggest that this modeling provides a framework for the description of the dynamics of stock markets intraday
price fluctuations.
\end{abstract}

\begin{keyword}
Non-extensive Statistical Mechanics \sep Stochastic Processes \sep Econophysics \sep Non-linear Dynamics \sep High-frequency Returns
\PACS 89.65.Gh \sep 02.50.-r \sep 05.10.Gg \sep 05.45.Tp         
\end{keyword}
\end{frontmatter}

\section{Introduction}

The empirical probability distribution functions (PDFs) of price fluctuations of financial indices for different
 markets have been reported in the econophysics literature in recent years \cite{Mantegna,Voit}. Although there has
  been a huge progress in the statistical description of these fluctuations, a complete and consistent description
   of the distributions and its dynamics in the high-frequency regime is still lacking.

It is well known that high-frequency financial time series such as foreign exchange rates or stock market returns
 are long-memory non-Gaussian processes. Such complex behavior calls for advanced theories in the field of
  statistical physics. Whence, we consider the recently proposed non-extensive statistical physics \cite{Swinney},
   a theory that generalizes the extensive classical statistical theory for strong dependent variables.
    A large number of studies \cite{Kozuki,Ausloo,Matsuba,Ramos} of financial markets have already employed 
    the non-extensive statistics in their analysis.

In this paper, we investigate the intraday dynamics of the BOVESPA index, the financial index of the Brazilian 
stock market, one of the largest emerging markets in the world. We model the distributions of the index price
 fluctuations by q-Gaussians, which are a class of stable distributions that emerge from the non-extensive
  approach, and where the parameter q measures the degree of the non-extensivity of the stochastic process. 

We find that for time horizons less than one hour, the probability distributions
 of price fluctuations are well fitted by q-Gaussians with damped exponential tails. A q-Gaussian scaling
  at the centre of the distributions with q*=1.75 is observed, which holds over these high frequency time scales
   due to the presence of strong correlations. The quasi-stationary q*-Gaussian regime breaks down due to
    the exponentially truncated tails and a new intraday correlated regime emerges at larger time scales,
     in which the q*-Gaussian central part of the distributions are consumed by the exponential tails. 

The dynamics in the short-time regime is in agreement with the prediction of a super diffusive q-Gaussian
 stationary process governed by a nonlinear Fokker-Planck equation (FPE) that models the correlated anomalous diffusion found for 
 high frequency price fluctuations. Therefore, we present a coherent description encompassing non-extensive 
 statistics and an evolution equation.

This paper is organized as follows. In section 2, we describe the empirical observations for the high-frequency
 BOVESPA index. In section 3, we present the non-extensive statistical theory. In sections 4 and 5, these
  observations are analyzed by using the non-extensive approach. In section 6, we present some concluding remarks.

\section{Empirical results}

Our results are based on analyzing BOVESPA index high-frequency 30-seconds interval data collected from
 November 2002 to June 2004, a period without major local market disturbances. This data consists of a set 
 of 352500 prices (index values) p(t) which allows a fairly statistical analysis. 

From this database, we select a complete set of non-overlapping price fluctuations (returns): 

\vspace{-0.5cm}
\begin{eqnarray}
x_\tau(t)=ln \ p(t+\tau)-ln \ p(t)
\label{Eq1}
\end{eqnarray} 

\vspace{-0.5cm}
occurring in $\tau$-intraday interval. We define the normalized returns as:

\vspace{-0.5cm}
\begin{eqnarray}
x^N_\tau(t)=(x_\tau(t)-\mu_\tau)/\sigma_\tau
\label{Eq2}
\end{eqnarray}  

\vspace{-0.5cm}
where $\mu_\tau$ and $\sigma_\tau$ are respectively the mean and the standard deviation of $x_\tau(t)$.

To characterize quantitatively the observed stochastic process, we measure the probability distribution
 $P(x_\tau^N)$ of index price fluctuations for $\tau$ ranging from 1 to 60 minutes. The number of data in each
  set decreases from the maximum value of 176250 ($\tau$ = 1 min) to the minimum value of 29375 ($\tau$ = 60 min).

\begin{figure}[h]
\centering  
\includegraphics*[width=0.65\linewidth]{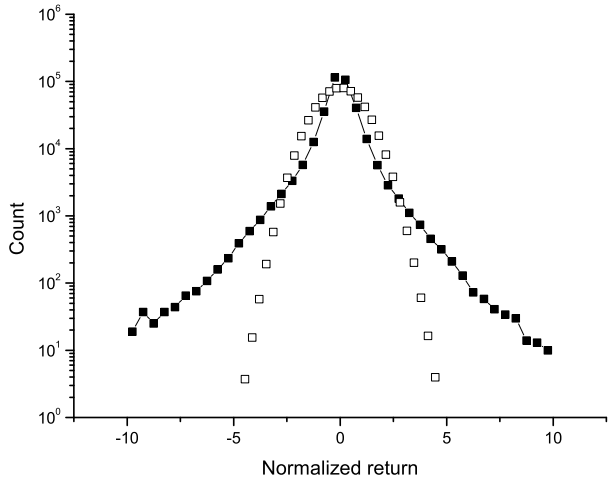}
\caption{Binned non-normalized frequency distribution of Bovespa index $\tau$ = 1 minute normalized returns in
logarithmic scale (full squares) and comparison with the expect distribution for a Gaussian process (open squares).}
\label{fig1}
\end{figure}

In figure 1 we show the measured (non-normalized) probability distribution for $\tau$ = 1 min normalized returns
 and compare with the expected distribution for the Gaussian process. The empirical distribution is clearly fat tailed,
 indicating that non-Gaussian time series analysis is required.

The scaling behavior of the distributions at coarser time scales has different regimes. At micro scales (typically
 shorter than one hour), correlations between successive price changes are dominant due to informational flow
  between the financial trades. On the other hand, at macro scales (typically larger than one month) macroeconomic
   rules govern the market drift and correspond to a Gaussian regime. Mesoscopic time scales can be associated
    with a transient regime between the microscopic and macroscopic ones. 

It is still an open question the evolution of the PDFs at micro and meso time scales. In this paper, we investigate
 the scaling behavior of the return distributions at microscopic time scales. As shown in figure 2, the data collapse
  into the   $\tau$ = 1 min distribution, but this stationarity breaks down at $\tau$ = 1 hour. This means that 
  the $\tau$ = 1 min return distribution functional form is preserved for several micro time scales, therefore
   exhibiting a short-time non-Gaussian intraday quasi-stationary regime.

\begin{figure*}[!htb]
   \centerline{
   \subfigure[]{\includegraphics[width=8.0cm]{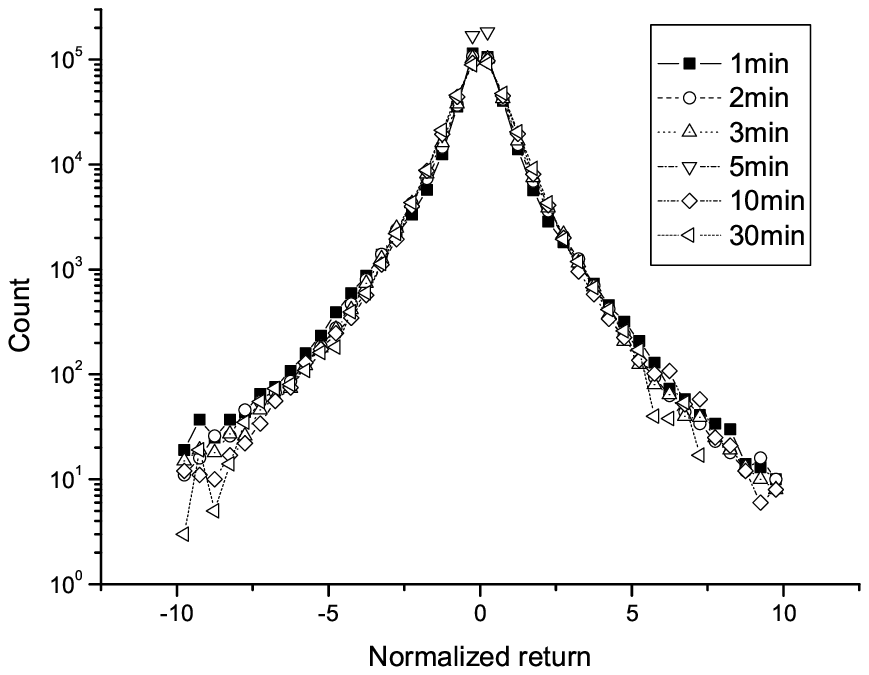}
   \label{}}
   \hfil
   \subfigure[]{\includegraphics[width=8.0cm]{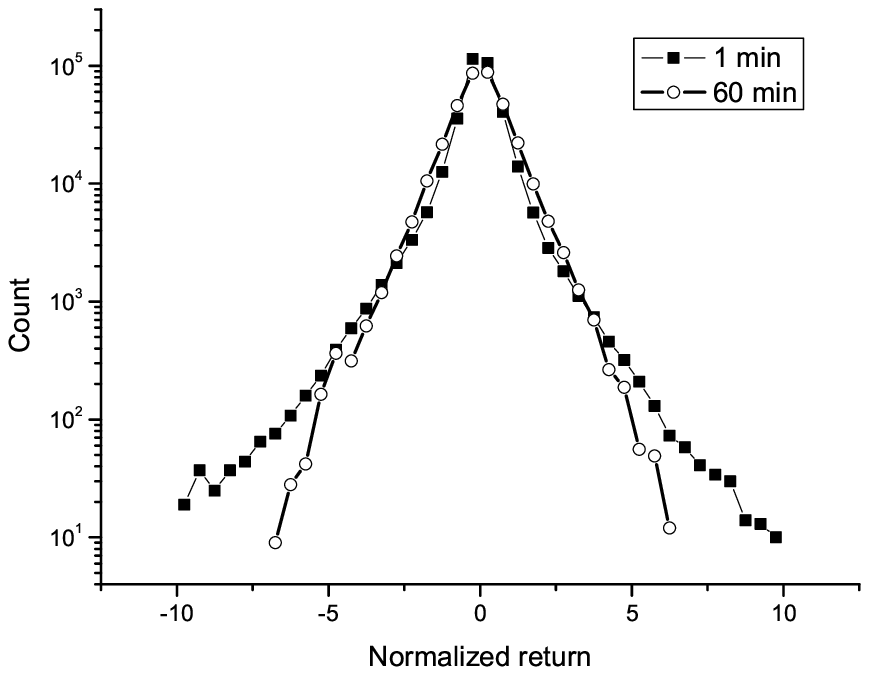}
   \label{}}
   }
   \caption{Distributions of normalized returns of the Bovespa index sampled with different time scales (shown in the inset) (a) the data collapses into a single distribution, (b) the collapse breaks down at $\tau$=60 min time scale.}
   \label{fig:2}
\end{figure*}

We also investigate the presence of linear and non-linear dependence between successive $\tau$ = 1 min normalized 
returns. In Figure 3(a) we show the linear autocorrelation $R(\Delta t)=<x_\tau^N(t)x_\tau^N(t+\Delta t)>$ according 
to the time lag $\Delta$t of the measured returns. Employing the standard least-squares fit for time intervals
 $\tau$ = 2-15 minutes, the semi-log plot indicates a short range exponential behavior $R(\Delta t) \sim 
 exp(-\Delta t/\Gamma)$ with characteristic time scale $\Gamma \cong$ 4 min. After $\Delta t \cong$ 20 min the 
 correlation is at the noise level.

\begin{figure*}[!htb]
   \centerline{
   \subfigure[]{\includegraphics[width=8.0cm]{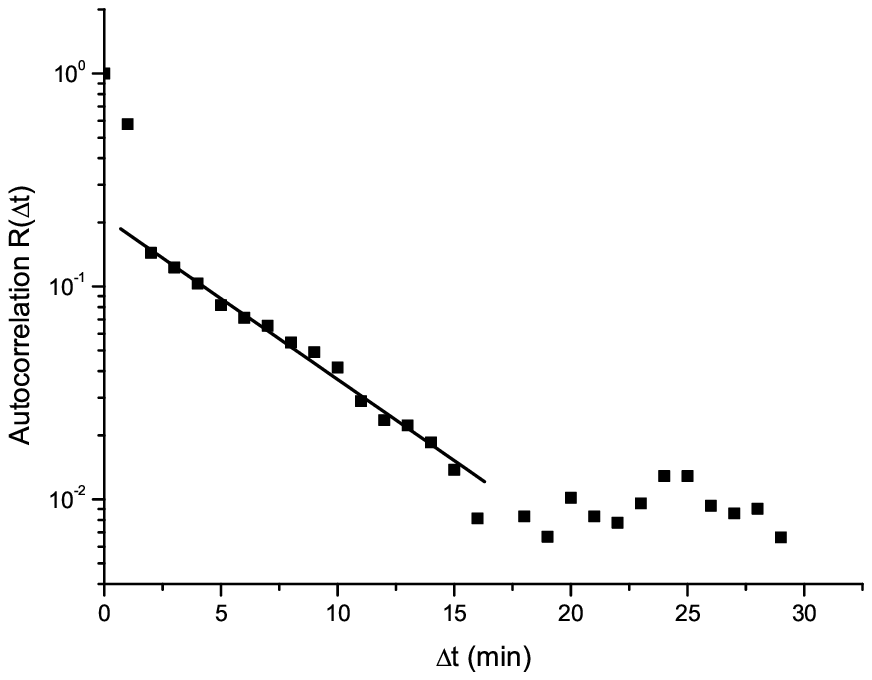}
   \label{}}
   \hfil
   \subfigure[]{\includegraphics[width=8.0cm]{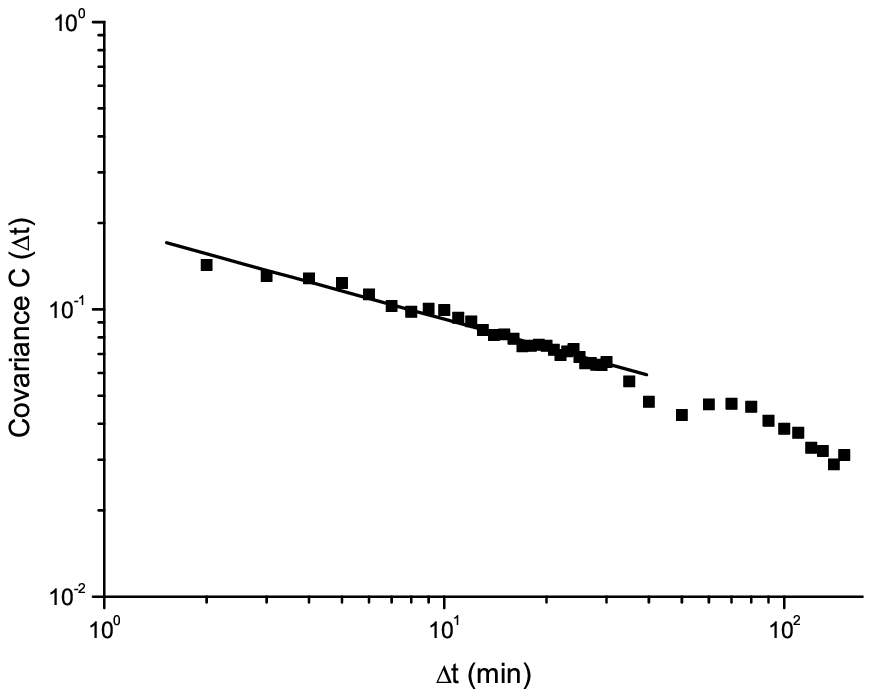}
   \label{}}
   }
   \caption{(a)  semi-log plot of the autocorrelation function of $\tau$ = 1 min normalized returns according to the temporal 
distance $\Delta$t of the measured returns.  The solid line represents the linear fit from $\Delta$t = 2 min to 15 min. 
The autocorrelation reaches noise level after $\Delta$t = 20 min.  (b)  log-log plot of the covariance function of the amplitude 
of $\tau$ = 1 min normalized returns according to temporal distance $\Delta$t of the measured returns.
  The solid line represents the linear fit from
 $\Delta$t = 2 min to 60 min.}
   \label{fig:3}
\end{figure*}
 \pagebreak

 In Figure 3(b) we show the covariance of the amplitude of the $\tau$ = 1 min normalized returns
 $C(\Delta t) = <|x_\tau^N(t)| \ |x_\tau^N(t+\Delta t)|> - <|x_\tau^N(t)|>^2$ according to lag $\Delta$t. Employing the
  standard least-squares fit for time intervals $\tau$ = 2-60 minutes, the log-log plot shows a power-law dependence 
   $C(\Delta t) \sim \Delta t^{-\eta}$ with $\eta \cong$ 0.3 along all intraday time lags. This result signals a
    long-memory process, without a measurable temporal scale of 
decay of correlation within the analyzed regime.  These findings are quantitatively similar to previous results 
for the U.S. market \cite{Liu}.

The anomalous diffusive process driven by the long and/or short range dependence of price fluctuations can be 
characterized by the scaling behavior $\sigma(\tau) \sim \tau^H$ of the standard deviation of the distributions. 
Through a log-log plot shown in figure 4 we estimate the scaling (Hurst) exponent H $\cong$ 0.7. This non-trivial
scaling exponent is the signature of a super diffusive behavior for the intraday time scales. Actually, it is a stylized 
fact that the intraday financial market generates strong dependence in the consecutive price formation.

\begin{figure}[h]
\centering  
\includegraphics*[width=0.65\linewidth]{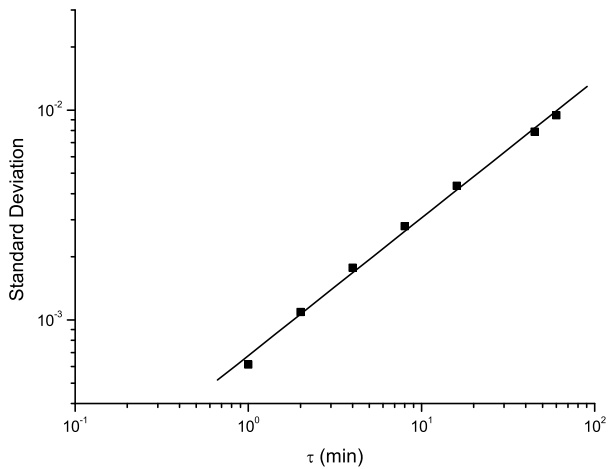}
\caption{Log-log plot of the standard deviation of the empirical return distributions as a function of time 
scale $\tau$. The solid line represents the linear fit from $\tau$ = 1 min to $\tau$ = 60 min.} \label{fig4}
\end{figure}

\section{Non-extensive statistics}

The quasi-stationarity property of the microscopic time scale distributions shown in figure 2 suggests the search 
for stable distributions to model them. Sums of independent or weak dependent stochastic variables lead to Gaussian
 or to L\'evy limit regimes \cite{Levy}. On the other hand, the presence of linear and non-linear dependence found
  for the one minute returns suggests the search for stationary distributions in a long memory environment.

Recently, a non-extensive statistical approach that generalizes the standard Boltzmann-Gibbs theory has been developed,
 as an attempt to describe non-equilibrium regimes and/or systems with strong dependence.

The cornerstone of this formalism is the entropy functional $S_q[\rho]=-\frac{1-\int[\rho(x)]^q dx}{1-q}$, 
where q $<$ 3 is a real parameter representing the degree of non-extensivity of the functional and $\rho$(x) is 
a normalized probability density function. The q $\rightarrow$ 1 limit recovers the classical extensive property.
 The optimization of $S_q$ under natural constraints leads to q-Gaussian PDFs defined as:

\begin{equation}
G_q(x,\tau) = \frac{1}{Z_q(\tau)} \left\{1-\beta(\tau)\left[(1-q)(x- \bar{x} (\tau))^2\right]\right\}_+{^\frac{1}{1-q}}
\label{Eq3}
\end{equation}

where $Z_q(\tau)$ is a normalization constant, $\beta(\tau)$ is a scale parameter ($\beta^{-1}$ is proportional to the variance
 of the distribution) and the subindex + indicates that  $G_q(x,\tau) = 0$ if the expression inside the brackets is
  non-positive. The usual Gaussian distribution is recovered in the $q\rightarrow1$ limit.

An important property of the q-Gaussian (3) is that, with appropriate time-dependent parameters $\beta(\tau)$ and 
$Z_q(\tau)$, it is the invariant solution for a class of (correlated) anomalous diffusive processes governed by 
non-linear Fokker-Planck equation of the form \cite{Bukman}:

\vspace{-0.5cm}
\begin{equation}
\frac{\partial P(x,\tau)}{\partial \tau}=-\frac{\partial}{\partial x}\left[F(x)P(x,\tau)\right]+D\frac{\partial^2}{\partial x^2}\left[P \ ^{2-q}(x,\tau)\right]
\label{Eq4}
\end{equation} 

\vspace{-0.1cm}
where $F(x) = a - b.x$ is a mean reverting force with $b\not=0$. The explicit time dependence of $\beta(\tau)$ and $Z_q(\tau)$ are:

\vspace{-0.4cm}
\begin{eqnarray}
&&\beta(\tau)^{\frac{-(3-q)}{2}}=\beta(\tau_0)^{\frac{-(3-q)}{2}}exp\left[-b(3-q)(\tau-\tau_0)\right]+ \nonumber \\
&&+2Db^{-1}(2-q)\left[\beta(\tau_0)Z^2_q(\tau_0)\right]^{\frac{q-1}{2}}\left\{1-exp\left[-b(3-q)(\tau-\tau_0)\right]\right\}
\label{Eq5}
\end{eqnarray}

\vspace{-0.8cm}
\begin{equation}
Z_q(\tau)=Z_q(\tau_0)\left\{(1-\Delta_q)exp\left[-(\tau-\tau_0)(b(3-q))\right]+\Delta_q\right\}^{\frac{1}{3-q}}
\label{Eq6}
\end{equation}

\vspace{-0.1cm}
where $\Delta_q=2b^{-1}(2-q)D\beta(\tau_0)Z_q^{q-1}(\tau_0)$. Equation (5) describes anomalous scaling of the inverse variance parameter $\beta$ for $q\not=1$. In the limit
 of weak reverting force and for time scales such that $b(3-q) << (\tau-\tau_0)^{-1} << 
 2D(2-q)(3-q)\beta(\tau_0)Z^{q-1}(\tau_0)$, one gets the result for the free diffusion process $(F(x)=0)$ \cite{Plastino}:

\vspace{-0.4cm}
\begin{equation}
\frac{\beta(\tau)}{\beta(\tau_0)}\approx (\tau - \tau_0)^{\frac{-2}{(3-q)}}
\label{Eq7}
\end{equation}

\vspace{-0.1cm}
Super diffusion occurs for q $>$ 1. This is the range of interest of the non-extensive parameter for
application to our intraday data. Hence, parameter q represents the degree of the resulting anomalous
diffusion from the underlying interaction among financial trades.

We model the normalized returns at time scale $\tau$ by the stochastic variable $x$ in Eq. (4). Thus, the 
solutions $G_q(x,\tau)$ describe the empirical distribution $P(x^N_\tau)$. 

The mean-reverting parameter \textit{a} controls the average equilibrium value of the stochastic variable
and does not affect the diffusion properties. In our application for normalized returns, \textit{a} is set 
equal to zero. On the other hand, from (5), the mean-reverting rate parameter \textit{b} is crucial for the 
attainment of a long-time equilibrium solution, but it is worth noting that, as \textit{b} gets smaller, the 
spread of the distributions holds over longer periods and the equilibrium solution may not be observable.

Substituting the explicit expression of $G_q(x,\tau)$ in (4) it can be shown \cite{Celia} that the stochastic
variables, whose addition has the q-Gaussian as a limit distribution, are characterized by non-null linear
correlation (for $b\not=0$) and non-null higher-order correlations (for $q\not=1$). In our application, these
step variables are the one minute normalized returns, which, according to the empirical findings, are consistent
with this modeling with $q\not=1$ and $b\not=0$, for microscopic time scales.

\section{Non-extensive modeling of the intraday return distributions}

It is apparent from the results sketched in the previous section that the q-Gaussians are good candidates 
for describing the anomalous diffusion of financial market indices. Since the correlations are stronger for shorter
 trading time intervals, we focus on the microscopic time scales less than one hour (1 min $< \tau <$ 60 min).

\begin{figure*}[!htb]
   \centerline{
   \subfigure[]{\includegraphics[width=8.0cm]{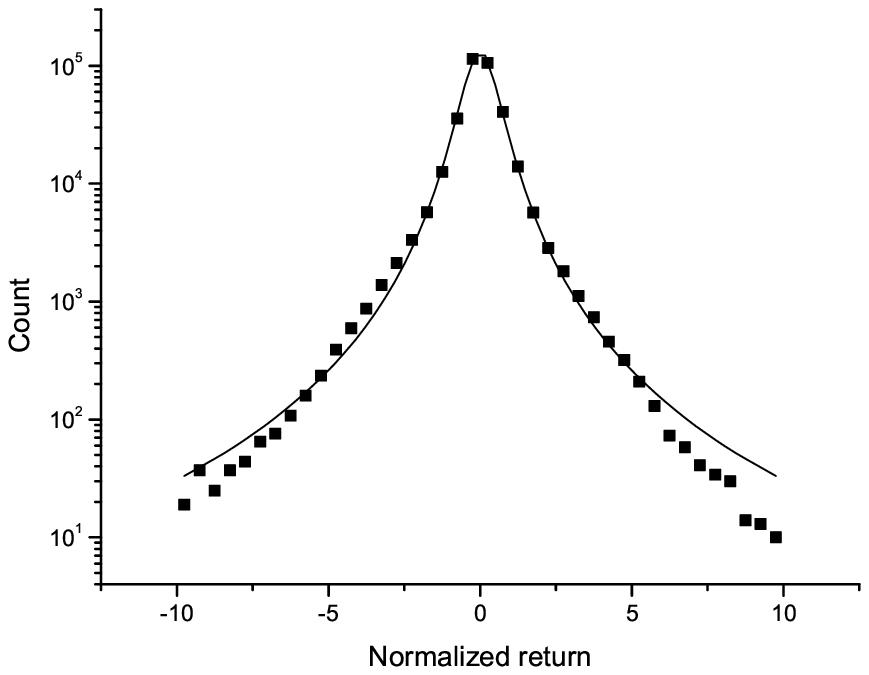}
   \label{}}
   \hfil
   \subfigure[]{\includegraphics[width=8.0cm]{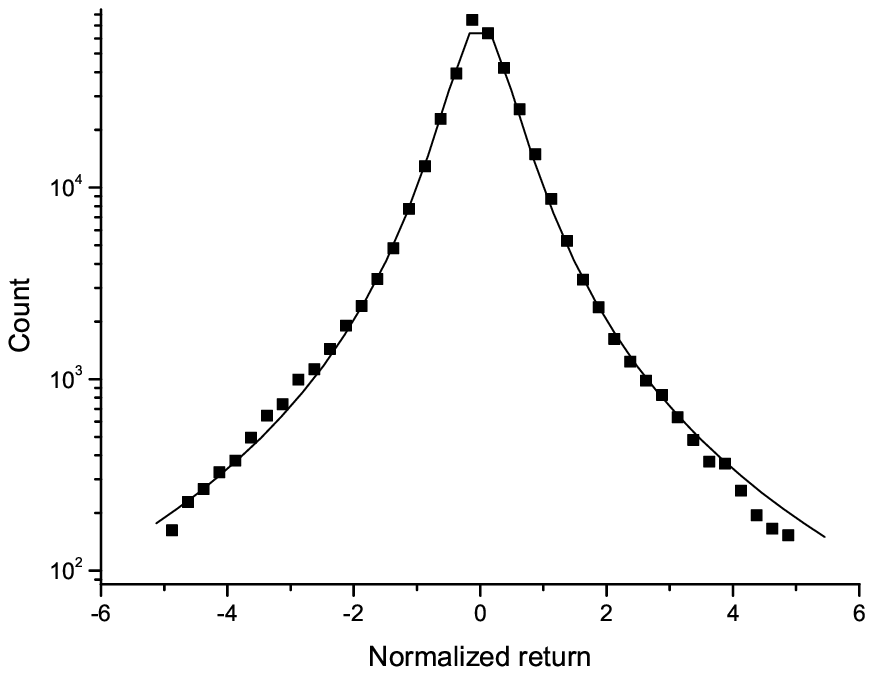}
   \label{}}
   }
   \caption{(a) Probability distribution function of normalized returns for time interval $\tau$ = 1 min: BOVESPA index data 
(black squares) and best fit of the data with Eq (3): (a) range $x_1^N  \in $ [-10,+10]; best fit q=1.64 and $\beta$=3.36. (b)  
range $x_1^N  \in $ [-5,+5]; best fit q*=1.75 and $\beta$=4.47.}
   \label{fig:5}
\end{figure*}

We determine the parameters of the model by minimizing the mean square deviation between the empirical distribution
$P(x^N_\tau)$ and the q-Gaussian distribution form $G_q(x,\tau)$. This method was applied throughout this paper. 
The free parameters are q and $\beta$ controlling respectively the shape and the width of the q-Gaussian.
In Figure 5(a) we plot the binned empirical distribution
of normalized $\tau$ = 1 min returns and the optimal fit. We find that the empirical distribution is well fitted 
with a q-Gaussian curve within the range $x_1^N \in$ [-5,+5] and it is overestimated outside 
this range. The optimal non-extensive parameters are $q=1.64$ and $\beta = 3.36\sigma^2_1$, where $\sigma_1=5.7 . 10^{-4}$.

We now search for the optimal fit only for the bulk of the $\tau$ = 1 min returns, defined as
returns with magnitude less than or equal 5$\sigma_1$. Figure 5(b) shows that a good agreement with the q-Gaussian profile 
is obtained when $| x^N_1(\tau)| < 5$ with $q*=1.75$ and $\beta = 4.47/\sigma^2_1$. 

Comparing the above results, we find that the inclusion of the tails of the empirical distributions in the fit has the
effect of diminish the optimal q-value. Nevertheless, the fit at the tails are not satisfactory. Then, for our
non-extensive modeling, we consider only the bulk of the distribution. 

To analyze the stability of the market through the elapse time of observation of our data, we estimate the local
parameter q and investigate if it is highly fluctuating. In figure 6, we show the time series of the local q
parameters considering the bulk of $\tau$ = 1 min returns for each month. From this figure, they
are roughly constant over the months and are consistent with the optimal value q*=1.75 for the entire data set.
Hence, it is possible to assume that this parameter is an invariant of the market characterizing the micro time scales.

\begin{figure}[h]
\centering  
\includegraphics*[width=0.67\linewidth]{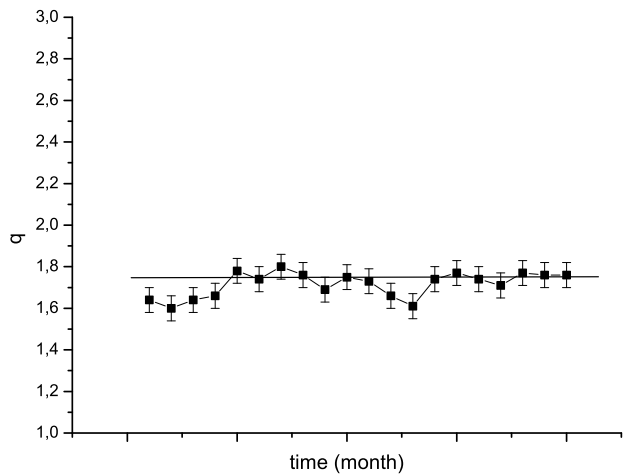}
\caption{Time series of optimal parameter q for the bulk of $\tau$ = 1 min returns collected in each month of the data set.
 The horizontal line represents the optimal parameter q*=1.75 for the whole period of the analyzed data.} \label{fig6}
\end{figure}

In figure 5(a), it was shown a clear deviation of the tails of the distribution from the q-Gaussian profile.
To investigate the tails, cumulative distributions $P_{cum}(x)$ of the positive and negative tails for normalized 
$\tau$ = 1 min returns are plotted in a semi-log space, as shown in figure 7. The linear dependence of the graph
suggests that the $\tau$ = 1 min returns exhibit an exponential decay in the tails. The straight lines are almost
parallel to each other, indicating that both tails follow approximately the same exponential law $P_{cum}(x) 
\sim exp(-(|x|-x_C)/\xi)$ with $\xi\cong4\sigma_1$.

\begin{figure}[h]
\centering  
\includegraphics*[width=0.67\linewidth]{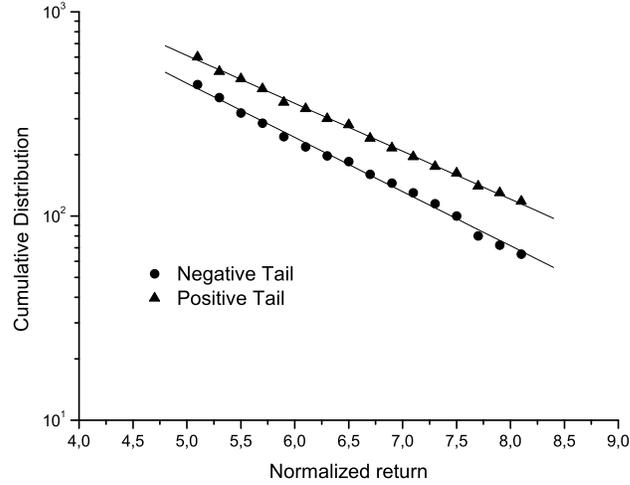}
\caption{Semi-log plot of the cumulative distribution of positive and negative tails for $\tau$ = 1 min normalized 
return distribution.} \label{fig7}
\end{figure}

This asymptotic exponential behavior for large returns also emerges as a robust characteristic of the intraday 
distributions at larger time scales, as shown in figure 8, where the distributions exhibit a tent-shape form, 
when plotted in a semi-log space.

\vspace{-0.2cm}
\begin{figure}[h]
\centering  
\includegraphics*[width=0.75\linewidth]{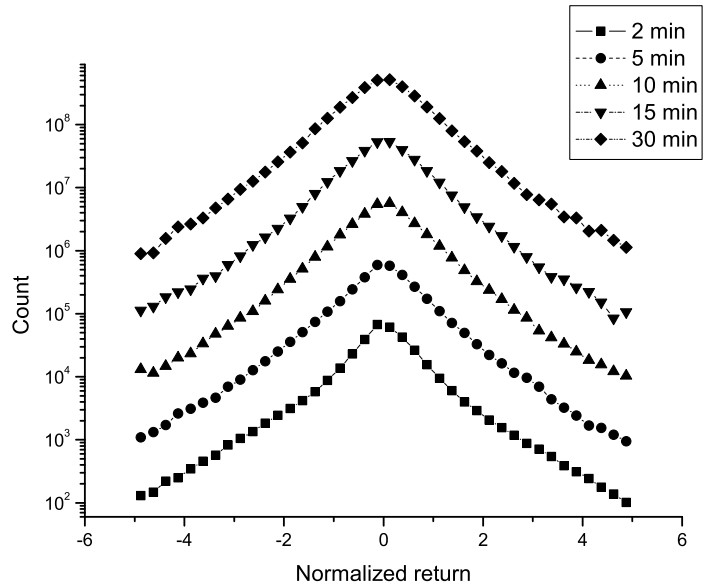}
\caption{Semi-log plot of the probability distribution of Bovespa index normalized returns in the range $x_1^N 
 \in $ [-5,+5]. For clarity, the distributions for successive $\tau$ are shifted up by a factor of 10 each. The inset
  shows the time scales.} \label{fig8}
\end{figure}

\vspace{-0.1cm}
The q-Gaussians have sharp peaks and power-law tails. For $q>5/3$, the q-Gaussian has an infinite second moment.
However, for large returns, the empirical distribution crosses over to simple exponential decay. The exponential
fall-off implies that the second moment is finite. These results suggest that the distributions
of high-frequency returns can be modeled by exponentially truncated q-Gaussian of the form:

\vspace{-0.8cm}
\begin{equation}
\left\{
\begin{array}{rcl}
P_q^T(x)= & CP_q(x), &\mbox{para } |x| \leq x_C\\[8mm]
P_q^T(x)= & CP_q(x)exp\left(-\frac{|x|-x_C}{\xi}\right), &\mbox{para } |x| > x_C\\
\end{array}\right.
\label{Eq8}
\end{equation}

In Figure 9 we show that Eq. (8) very well reproduces the distribution of $\tau$ = 1 min normalized returns
 of the Bovespa index with the fitting parameters $q*=1.75, \beta=4.47/\sigma^2_1,\xi=4\sigma_1$ and
$x_C=4\sigma_1$. Note that the value of $x_C$ is consistent with the region $|x_1|>5\sigma_1$ considered for the tails.

\vspace{-0.3cm}
\begin{figure}[h]
\centering  
\includegraphics*[width=0.65\linewidth]{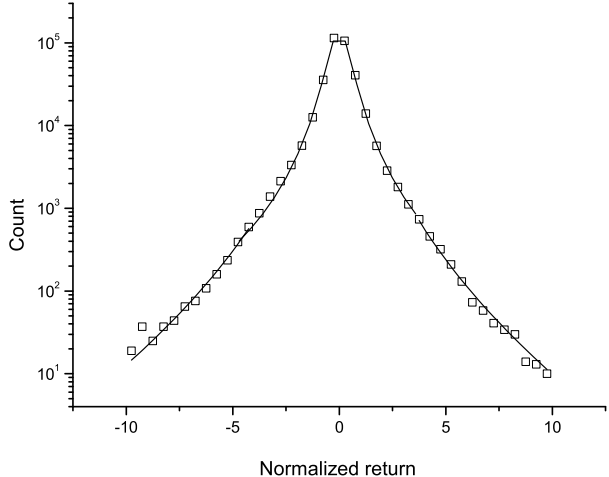}
\caption{Semi-log plot of the $\tau$ = 1 min normalized probability distribution (open squares) and the
 best-fit using an exponentially truncated q-Gaussian given by Eq. (8) with parameters q* = 1.75, $\beta$=4.47, 
 $\xi$ = 4 and $x_C = 4$.} \label{fig9}
\end{figure}

In section 2 it was shown through Figures 3-5 the presence of significant correlations of returns at microscopic 
time lags. It is also important to test the null hypothesis that the observed correlations cause the persistent fat tails 
of the high frequency distributions. In that case, destroying all correlations would almost vanish the fat tails for 
larger time scales. With this aim, we randomized the empirical series of $\tau = 1$ min returns, by shuffling them, 
creating an artificial series. 
In table 1, we show a comparison between the effective best-fit parameter q for the empirical (EMP) distributions and for 
distributions of the $\tau$-scale returns generated by the aggregation of the $\tau = 1$ min shuffled returns (SHU). While the randomized data have reached the Gaussian regime (q=1) at microscopic 
time scales, this regime is far from been reached by the real data in the intraday time scales. This result supports that the 
scaling property observed in figure 2(a) and the associated slow convergence to the stable Gaussian distribution 
are caused by the nontrivial correlations in the market data.

\vspace{-0.1cm}
\begin{table}[h]
\begin{center}
\begin{tabular}{|c|c|c|}
\hline
Time scale $\tau$ (min)	&\hspace*{4mm}EMP\hspace{4mm}	&\hspace*{4mm}SHU\hspace{4mm} \\[0.1mm]
\hline
1		&1.75	&1.75	\\[0.1mm]
\hline
2		&1.75	&1.65	\\[0.1mm]
\hline
4		&1.73	&1.50	\\[0.1mm]
\hline
8		&1.64	&1.33	\\[0.1mm]
\hline
16		&1.52	&1.24	\\[0.1mm]
\hline
30		&1.51	&1.18	\\[0.1mm]
\hline
45		&1.49	&1.07	\\[0.1mm]
\hline
60		&1.48	&1.06	\\[0.1mm]
\hline
\end{tabular}
\caption{Effective q-parameters for return distributions at several microscopic time scales, for the original empirical
$\tau$ = 1 min returns (EMP) and for the artificial shuffled $\tau$ = 1 min returns (SHU). The effective parameter q 
was adjusted for $|x_\tau| < 5\sigma_\tau$, which includes partially the tails of the distributions when $\tau > 1$ min.}
\end{center}
\end{table}

\section{Time evolution of the intraday return distributions}

In this section, we investigate the stationarity of the central part of the return distributions at microscopic
 time scales, shown in figure 2(a). We now consider the empirical distributions of rescaled returns, this is, 
 detrended original returns in units of $\sigma_1$:

\vspace{-0.5cm}
\begin{eqnarray}
x^R_{\tau}(t)=(x_\tau(t)-\mu_{\tau})/\sigma_1
\label{Eq9}
\end{eqnarray}

\vspace{-0.5cm}
We consider the $\tau$ = 1 min optimal non-extensive parameter q*=1.75 and obtain the best-fit parameter 
$\beta(\tau)$ for the central part of the rescaled return distributions at coarser time scales. In figure 10, 
we plot the distributions of figure 8 in rescaled variables. It is shown that the optimal q*-Gaussians well 
reproduce the bulk of the $\tau$-min rescaled return distributions in the microscopic regime.

\begin{figure*}[!htb]
   \centerline{
   \subfigure[]{\includegraphics[width=7.5cm]{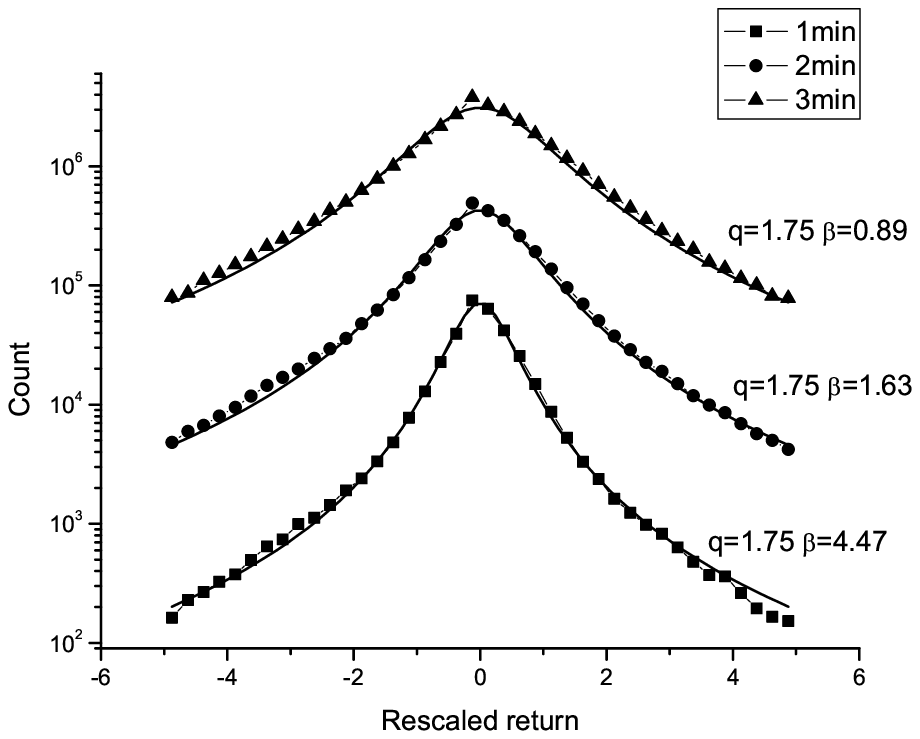}
   \label{}}
   \hfil
   \subfigure[]{\includegraphics[width=7.5cm]{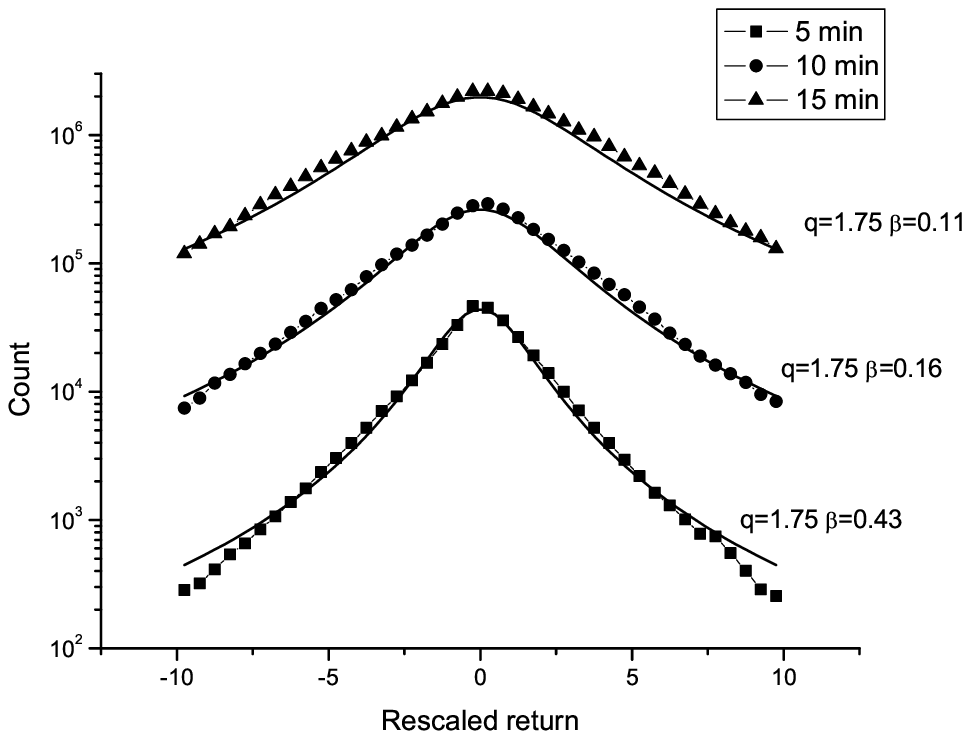}
   \label{}}
   }
   \caption{Rescaled return distribution for microscopic time scales and optimal q*-Gaussian fit with 
   parameter q* =1.75 : (a)   $\tau$ = 1, 2, 3 min  (b) $\tau$ = 5, 10, 15 min. Distributions are shifted up 
   by a factor 10 each, for best visualization. }   \label{fig:10}
\end{figure*}

The results presented so far lead to the hypothesis of a quasi-stationary q-Gaussian regime for microscopic
 time horizon. The presence of the exponential tails do not affect the q-Gaussian scaling at the centre of the 
 distributions, which holds over the microscopic time scales due to the presence of strong correlations. 
 
We now investigate the super diffusive time scaling behavior of the empirical distributions along the same lines of
\cite{Michael}. We test the observed behavior with the time evolution predicted by the non-linear Fokker Planck Eq. (4)
with q*=1.75.

\begin{figure}[h]
\centering  
\includegraphics*[width=0.65\linewidth]{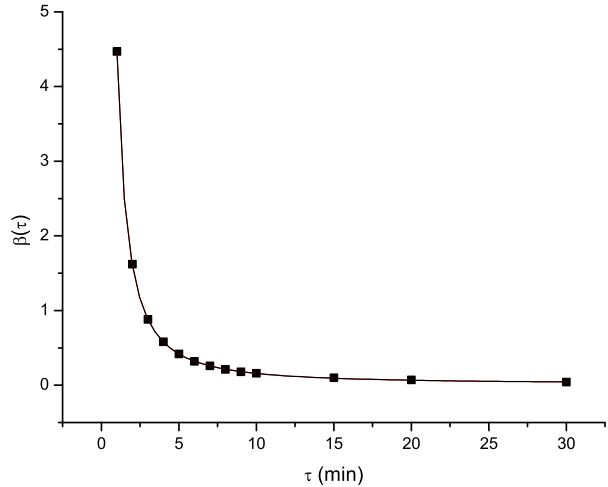}
\caption{Time evolution of parameter $\beta$ as a function of time scale $\tau$: Bovespa index data (black squares)
 and best fit (line) using Eq. (5) with q*=1.75 and $\tau_0$= 1 min. The estimated model parameters are
 D=0.0286 $\pm$ 0.002 and b= 0.025 $\pm$ 0.003.} 
\label{fig11}
\end{figure}

In figure 11, we model $\beta(\tau)$ with Eq. (5), where q=q* and $\tau_0$= 1 min. There are two free coefficients left,
namely the diffusion constant D and the mean-reversion rate b. We show that the empirical values can be well-fitted
by the theoretical values, for particular estimated coefficients $D=0.0286 \pm 0.002$ and $b= 0.025 \pm 0.003$. 
These numerical values are similar to previous results for the U.S. market \cite{Michael}. Therefore, our results for 
the intraday Brazilian market indicate a time evolution according to the nonlinear Fokker-Planck form.

\section{Discussion}

The data collapse shown in figure 2(a) reveals the existence of an invariant time scaling master curve for the 
description of high-frequency returns. The observation of invariant quantities in the economic systems, 
as in the physical systems, denotes the existence of conservation laws or same rules governing these processes.

Moreover, it is known that there is strong time dependence among the tick-by-tick price formation by the financial 
market traders. In this work, we model the high-frequency returns of BOVESPA index by considering the q-Gaussian
distributions, which are stationary solutions for a class of anomalous diffusive processes driven by strong correlations. 

Adjusting only the central part of the one minute distribution with the q-Gaussian distribution function, we
obtain the optimal non-extensive parameter q*=1.75. This value is remarkably constant over the two-year period
of our data, as shown in figure 6. Similar values (q=1.77 and q=2) were obtained for the central part of the
return distributions for the Japanese \cite{Matsuba} and Korean \cite{Lee} stock indices, respectively. On
the other hand, this value is bigger than previous ones reported in the literature for the U.S. market
\cite{Osorio}, which were obtained adjusting the whole range of the empirical returns. Actually, as shown in 
Figure 5, the inclusion of the damped tails of the distributions in the fit has the effect of diminish the optimal 
q value, but the observed exponential fall-off behavior of our empirical data disable the modeling with
q-Gaussians at the tails, due to their power-law character.

Figure 9 shows that Eq. (8) very well reproduces the distribution of $\tau$ = 1 min normalized returns 
of the BOVESPA index. Exponential tails in the foreign exchange and stock return distributions have been reported 
for some world markets \cite{Bouchaud,Silva,Miranda,Matia,Tang} for micro and meso time scales, although several
authors have reported power-law decays \cite{Plerou,Gorski,Mizuno}, including those employing the non-extensive
approach \cite{Kozuki,Ausloo,Ramos}. Our empirical distributions consist of a significant number of data which allows a 
fairly statistical analysis including the tails. The exponential tails may due to finite-size effects of
the financial market, which prevents the development of scaling invariant power-law tails. This effect is expected to be
stronger in the case of emerging markets. This result is also in according to previous findings
for the daily return distribution tail of the BOVESPA index \cite{Miranda}. 

We have shown that the index time series in the high frequency regime is a long memory process where 
the Hurst exponent is significantly greater than 1/2. This long-memory leads to stationary leptokurtic
distributions with the observed scaling behavior of $\beta(\tau)$ for microscopic time scales. The fast decay of 
$\beta(\tau)$ signals this super-diffusive character of the market dynamics.

The anomalous scaling (5) with two free parameters, D and b, reasonably fits the time evolution of the PDFs at micro scales. 
This modeling is able to predict the super diffusive behavior of this quasi-stationary regime. The estimated 
value for parameter \textit{b} is small, but significantly different from zero. This result is consistent with 
non-null linear correlation of normalized returns for time lags $\Delta t < 20$ minutes shown in figure 3(a) and a
lifetime of the regime towards the limit distribution that encompass the microscopic time scales. Recent 
analysis of the S$\&$P500 signal \cite{Ausloo} also shown that the first Kramers-Moyal coefficient of the FPE 
is almost zero, indicating almost no restoring force. The stable non-extensive parameter $q*>1$ in Eq. (4) implies 
the existence of higher-order correlations \cite{Celia}, modeling
the non-linear memory effects illustrated in figure 3(b). Hence, we conclude that the non-linear Fokker-Planck equation (4)
captures the main features of the dynamical evolution of stock price returns at this time horizon.

Another dynamical foundation of non-extensive statistics have been applied to financial indices \cite{Kozuki,Ausloo}, 
in which the price fluctuations are described by Brownian motion subject to a power-law restoring force and a 
Gaussian white noise with a slow varying amplitude. This leads to linear FPE for temporal evolution of the 
distribution function with time-dependent diffusive coefficient. However, if we take into account the strong 
correlations of the price changes at microscopic time scales, the description by means of a non-linear FPE seems 
to be more appropriate.

The q-Gaussian regime breaks down due to the exponentially truncated tails that prevent the attainment of the 
equilibrium solution. On the other hand, the long-range correlations exhibited by the intraday diffusive process 
further delay the convergence to the long time Gaussian regime, giving rise to a new intermediate regime at
mesoscopic time scales, in which the q-Gaussian central part of the distributions are consumed by the exponential
tails. This result is in accordance with previous findings \cite{Silva} for probability distributions of stock 
returns at mesoscopic time horizons that follow an exponential function.

In conclusion, we found that the intraday stock price fluctuations in Brazil have an intermediate behavior between
the non-extensive q-Gaussian and the extensive Gaussian regimes characterized by $\tau$-dependent exponential tails of
the return distributions. We have shown that a model based on a non-extensive approach that captures the strong
dependence of the stochastic variable robustly accounts for the observed time evolution of the return distributions
at micro time horizons and reproduces the crossover between micro and meso time lags. This exponentially damped, 
non-extensive, dynamical behavior should provide a framework to investigate other high-frequency stock market
time series.

\section{Acknoledgements}

This work is supported by the Brazilian agencies CAPES and CNPq.

\end{document}